\documentclass[10pt,showpacs,amsmath,amssymb,floatfix,superscriptaddress,longbibliography,twocolumn,eqsecnum]{revtex4-2}
\usepackage{amsmath}
\usepackage{physics}
\usepackage{graphicx}
\usepackage[usenames,dvipsnames,svgnames,table]{xcolor}
\usepackage{tcolorbox}
\usepackage{booktabs}
\usepackage{makecell} 
\usepackage{tabularx}
\usepackage{chngcntr}
\usepackage[utf8]{inputenc}
\counterwithout{equation}{section}
\counterwithout{figure}{section}
\usepackage[unicode=true,
            pdfusetitle,
            bookmarks=true,
            bookmarksnumbered=false,
            bookmarksopen=false,
            breaklinks=true,
            pdfborder={0 0 0},
            backref=false,
            colorlinks=true,
            hypertexnames=false]{hyperref}
\hypersetup{linkcolor=NavyBlue,urlcolor=NavyBlue,citecolor=NavyBlue}

\begin{document}
	
\title{Entanglement dynamics of multi-fluxonium-qubits under Non-Markovian TLS noise}
\author{Chenghong Ji}
\affiliation{College of physics, Hangzhou Dianzi University, Hangzhou 310018, China}
\author{Chaoying Zhao}
\email{zchy49@163.com}
\affiliation{College of physics, Hangzhou Dianzi University, Hangzhou 310018, China}
\affiliation{State Key Laboratory of Quantum Optics Technologies and Devices, Shanxi University, Taiyuan, 030006, China}
\affiliation{ Zhejiang Key Laboratory of Quantum State Control and Optical Field Manipulation, Hangzhou Dianzi University, Hangzhou, 310018, China}
\date{\today}

\begin{abstract}
The research on open quantum systems is important for both quantum computing and quantum sensing. So far, we can only use the main equation to make an approximate description. The dynamics of a single Fluxonium qubit under Markovian environment satisfied Lindblad Master Equation. In experiments,  pulse sequence dynamic decoupling (DD) can enhance the coherence of qubits and effectively suppress noise. Two Fluxonium qubits sensitive to two-level systems (TLS) noise. TLS formed by material defects results in noise with significant non-Markovian characteristics. The dynamics of non-Markovian noise satisfied the post Markov Master Equation (PMME). The TLS noise spectrum is mainly concentrated in low frequencies, so traditional DD cannot effectively suppress TLS noise. The relaxation and dephasing behavior with a complex dynamic characteristics. Based on Ornstein-Uhlenbeck process, we put forward a novel DD sequence and design a TLS-tailored dynamical decoupling protocol by optimizing pulse locations to minimize noise power spectral overlap with the Lorentzian shape. Using PMME-consistent framework, we can obtain a stronger low-frequency suppression and significantly prolong both Bell-based fidelity and entanglement. We explore specific DD design and precise modeling of entanglement dynamics under non-Markovian TLS noise. Our dynamical decoupling protocol can effectively improve entanglement gates fidelity in NISQ quantum devices.
\end{abstract}

\maketitle

\section{Introduction}

Superconducting quantum circuits represent one of the core platforms for quantum computing \cite{1,2}, with Fluxonium qubits garnering significant attention due to their low-frequency operation (0.1–1 GHz) and long coherence times. Compared to transmon qubits, Fluxonium qubits exhibit robustness against high-frequency noise through Josephson junction chains and high-inductance designs \cite{3,4,5,6,7,8}. However, they remain sensitive to low-frequency noise, particularly defects such as two-level systems (TLS) in amorphous media \cite{9,10,11}. TLS defects induce non-Markovian noise through electric field or flux coupling, manifesting as time-dependent memory effects that cause oscillatory fidelity decay and rapid loss of entangled states (e.g., Bell states), especially in two-qubit systems \cite{11,12,13,14,15,16}.

The millisecond scale slow switching dynamics of non-Markovian noise violate the Markov approximation, rendering traditional Lindblad master equations inapplicable \cite{12,13,14}. The Post-Markovian Master Equation (PMME) captures memory recirculation effects by introducing time-nonlocal dissipation terms, enabling the reversal of decoherence \cite{17,18}. However, existing studies predominantly focus on single-qubit relaxation, lacking systematic analysis of two-qubit Fluxonium entanglement dynamics. To actively mitigate decoherence induced by environmental noise, dynamic decoupling (DD) \cite{19} techniques were proposed, which suppress noise through pulse sequences and have proven successful under Markovian noise (CPMG sequences), but require optimized designs for the low-frequency TLS noise (Lorentzian spectral density) of Fluxonium.

 Odeh et al. validated TLS (density ~$10^{18}/m^3$) as dominant in Fluxonium through non-Markov relaxation spectroscopy, revealing memory recovery temporarily restores single-qubit coherence \cite{20}. Chen et al. mitigated TLS noise through phonon engineering, emphasizing its low-frequency characteristics \cite{21}. Nakamura et al. found non-Markovian noise accelerates two-qubit entanglement decay, with concurrence losing 80\% within microseconds \cite{22}. Regarding DD optimization, Nakamura et al. noted that conventional sequences (e.g., CPMG) become less efficient under slow TLS switching \cite{23}, while Ezzell et al. tested Heisenberg-Weyl group-based DDs demonstrating robustness to low-frequency noise but lacking TLS-specific optimization \cite{24,25}. We find that precise modeling of entanglement dynamics and specific DD design under TLS noise in two-qubit Fluxonium systems remain under-explored, limiting entanglement gate fidelity improvements in NISQ devices.

In this paper, we propose a theoretical framework to study fidelity and entanglement evolution in two-Fluxonium-qubits under non-Markovian TLS noise, and design TLS-optimized DD protocols. We employ PMME to model quantum-bit interactions with the TLS bath, combining Lorentzian spectral density (coupling strength 10–100 kHz) to quantify fidelity and concurrence evolution. This reveals complex memory effects on entanglement ($20\%$ states). We propose a DD sequence based on the Heisenberg-Weyl group, optimizing pulse intervals to significantly suppress decoherence. Numerical results demonstrate entanglement lifetimes extended to 10 times that of conventional DD, with fidelity approach to $99.9\%$, providing robust assurance for NISQ devices.

We establish a theoretical-experimental bridge by integrating PMME with experimental parameters (TLS density $10^{18}/m^3$). We systematic analysis entanglement dynamics in two-Fluxonium-qubits and propose a TLS-specific DD protocol that outperforms generic sequences. These will help our deepen understanding of non-Markovian noise and guide the application of Fluxonium in NISQ devices.

The paper is structured as follows: Section I introduces the two-Fluxonium-qubits-TLS Hamiltonian and PMME framework, and then derive the dynamical equations and analyzes fidelity and entanglement dynamics without DD sequences, and finally describes the TLS-optimized DD protocol and its performance. Section II presents numerical results and test DD sequences on the IBM Q platform. Section III makes a conclusion.

\section{THE THEORETICAL MODEL}

In superconducting quantum circuit, the fluxonium qubit combining a chain of Josephson junctions with a large inductance, operating at frequencies in the range of $0.1-1GHz$ and offering several hundred micro-seconds long coherence time. Two fluxonium qubits can couple with capacitance or inductance, and Hamiltonian including single-qubit, inter-qubit coupling, and interactions with a bath of two-level-system (TLS) noise.

\begin{figure}[htbp]
    \centering
\includegraphics[width=0.9\linewidth]{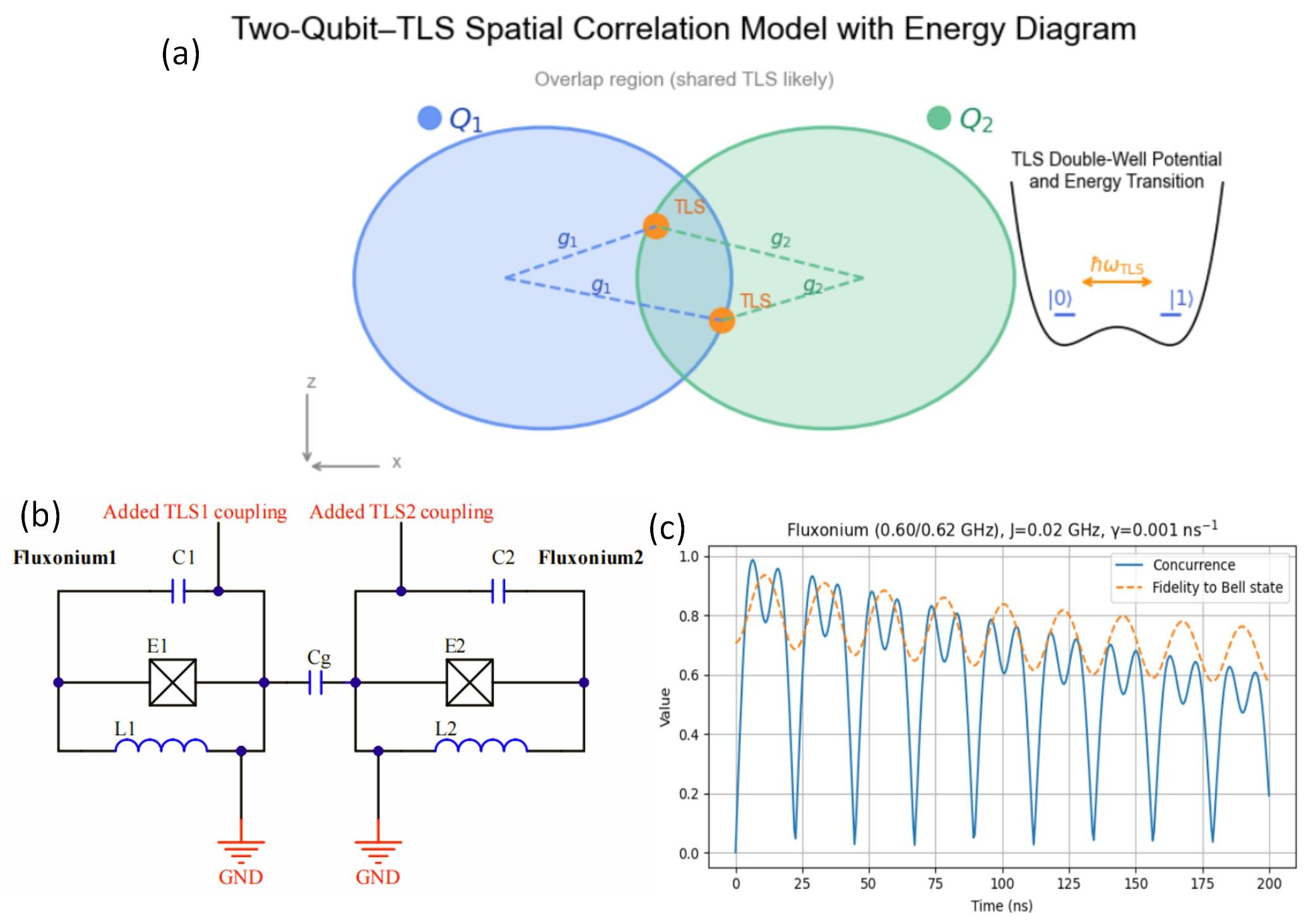}
    \caption{(a)Two fluxonium qubits circuit schematic: two fluxonium qubits (Josephson junctions $E_1/L_1$ and $E_2/L_2$) interact via the coupling capacitor $C_g$; each qubit is capacitively coupled to its local TLS through $C_1$ and $C_2$ (labeled “Added TLS coupling” in red line). $E_1$ and $E_2$ are Josephson junctions, $L_1$ and $L_2$ are super-inductors; GND denotes the common ground. (b)Time evolution of two-qubit entanglement and Bell-state fidelity. Initial state $|10\rangle$. Blue line: concurrence. Orange dashed line: fidelity with the Bell based $(|01\rangle + |10\rangle)/\sqrt{2}$. The curves show swap oscillations damped by dephasing.}
\end{figure}

We consider two fluxonium qubits capacitively coupled with a small inter-node capacitor $C_g$, as shown in Fig.1(a). Each fluxonium consists of a Josephson element shunted by a super-inductance $L_{1,2}$ and a capacitance $C_{1,2}$ to ground. In the qubit subspace, this capacitive channel realizes a predominantly exchange–type interaction (with a possible weak longitudinal component). Retaining only the qubits and the capacitive coupling, the total Hamiltonian reads (see  Supplemental Material)
\begin{equation}
	\resizebox{\linewidth}{!}{$
		\begin{aligned}
		\hat{H}_{\mathrm{tot}}
		&= 4E_{C_1}\hat{n}_{1}^{2} + 4E_{C_2}\hat{n}_{2}^{2} + 8E_{C_{12}}\hat{n}_{1}\hat{n}_{2}
		+ \sum_{i=1}^{2}\!\left[\frac{(\hat{\Phi}_i-\Phi_{\mathrm{ext},i})^{2}}{2L_i}
		- E_{J,i}\cos\!\left(\frac{2\pi}{\Phi_0}\hat{\Phi}_i\right)\right] \\
		&\quad + \sum_{i=1}^{2}\frac{\hbar\Omega_i}{2}\hat{\tau}^{(i)}_{z}
		+ \sum_{i=1}^{2}\lambda\!\left(\cos\theta_i\,\hat\tau^{(i)}_{x}+\sin\theta_i\,\hat\tau^{(i)}_{z}\right)(2e\,\hat{n}_i)
		+ \sum_{i=1}^{2} f_i(t)(2e\,\hat{n}_i)
		\end{aligned}
		$}
\end{equation}

where $i=1,2$ label the two fluxonium nodes, $\hat n_i$ is the Cooper-pair number and $\hat\Phi_i$ is node-flux operators with $[\hat\Phi_i,\hat n_j]=i\hbar\,\delta_{ij}$, and $\Phi_0=h/2e$ is the flux quantum. The self/mutual charging energies are $4E_{C_1}\hat n_1^2+4E_{C_2}\hat n_2^2+8E_{C_{12}}\hat n_1\hat n_2$, the cross term encodes capacitive inter-qubit coupling.  The inductive energy of the super-inductor is $(\hat\Phi_i-\Phi_{\mathrm{ext},i})^2/(2L_i)$ and  the Josephson potential is $-E_{J i}\cos\!\big(2\pi \hat\Phi_i/\Phi_0\big)$. Each node hosts a two-level system defect (TLS) with splitting $\hbar\Omega_i$,  $\hat\tau^{(i)}_{x,z}$ are Pauli operators and $\theta_i$ is the mixing angle. The interaction $\lambda\big(\cos\theta_i\,\hat\tau^{(i)}_x+\sin\theta_i\,\hat\tau^{(i)}_z\big)(2e\hat n_i)$ describes charge–dipole coupling, and $f_i(t)(2e\hat n_i)$ denotes a classical microwave driven on node $i$. The two fluxonium qubits are biased at zero-flux sweet spot $\Phi_{ext}/\Phi_0=0$ to suppress first-order flux noise, and without microwave driven is applied. The two qubits interact capacitively through the coupler $C_g$, which yields an exchange rate $J$. We solve the Lindblad master equation with pure-dephasing collapse operators. $\sqrt{\gamma}\, \hat{\sigma}_z\,\!\otimes\!\,\hat{I}$ and $\sqrt{\gamma}\,\hat{I}\,\!\otimes\!\,\hat{\sigma}_z$ by using \texttt{QuTiP}'s \texttt{mesolve}. 

We set parameters: $f_{q_1}=0.60GHz$, $f_{q_2}=0.62GHz$ , detuning $\delta=f_{q_1}-f_{q_2}=20MHz$, the coupling $J=0.02GHz$ is typical for a capacitive link, and the dephasing rate $\gamma=0.001ns^{-1}$ corresponds to $T_\phi\!\approx\!1/\gamma=1\mu s$, giving a visible but moderate damping envelope. All frequencies are implemented as angular frequencies $\omega=2\pi f$ (in $ns^{-1}$), whereas rate $\gamma$ is not multiplied by $2\pi$. This near-resonant choice ($\delta\ll f_{q_i}$) produces clear swap oscillations within  $t\in[0,200ns]$ window and avoiding the idealized $\delta=0$ limiting, and the exchange frequency with splitting $\Omega=\sqrt{\delta^2+4J^2}$. We set $\hbar=1$, the initial state is $\ket{10}$. The concurrence and the fidelity with Bell based $(\ket{01}+\ket{10})/\sqrt{2}$ are evaluated over time, as shown in Fig. 1(b). In order to avoiding crossing occurs between $\ket{10}$ and $\ket{01}$, we adopt half-swap, which happens near $t_{swap}\approx \pi/2\Omega$, and the oscillations are damped by the dephasing rate $\gamma$. Two qubits undergo pure dephasing due to ensembles of TLS fluctuators, populations are conserved while coherences decay.

\begin{figure*}[!htbp]
	\centering
\includegraphics[width=\textwidth]{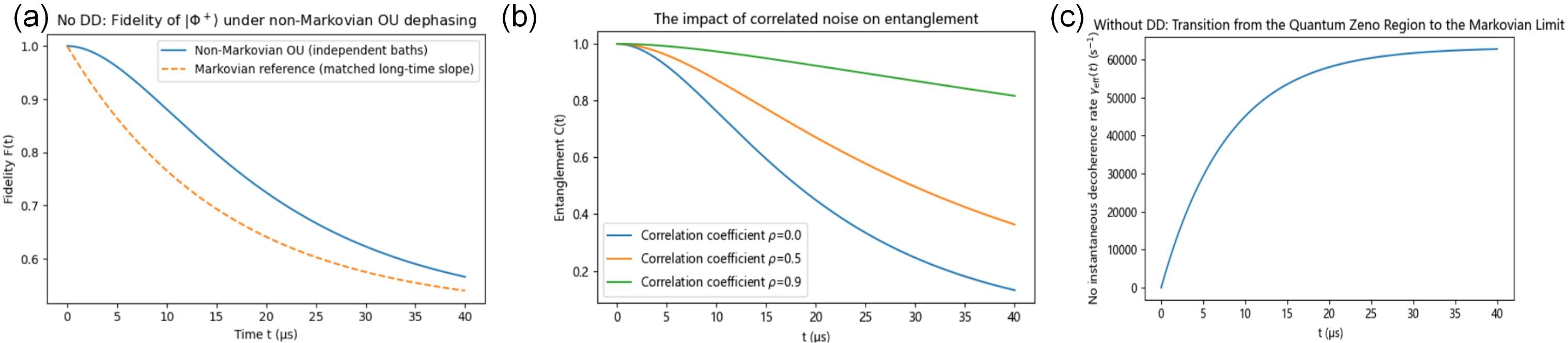}
	\caption{{No-DD}-(a) Under OU non-Markovian dephasing, the fidelity of \(|\hat{\Phi}^+\rangle\) decays more slowly than the Markovian exponential; (b) increasing in noise correlation \(\rho\) slows down the concurrence \(C(t)\) decay; (c) the effective dephasing rate \(\gamma_{eff}(t)=-\,d\ln F/dt\) rises from near zero and saturates around \(t\!\sim\!\tau_c\), indicating a Zeno-to-Markov crossover.}
\end{figure*}

In interaction representation, Hamiltonian with respect to the qubit splittings is \cite{24}
\begin{equation}
\hat{H}(t)=\frac{1}{2}\sum_{i=1}^2 \xi_i(t)\,y_i(t)\,\hat{\sigma}_z^{(i)}
\end{equation}
where $\xi_i(t)$ is the stochastic frequency shift acting on qubit $i$, and $y_i(t)\in\{\pm1\}$ is the modulation function; in the no-DD baseline, $y_i(t) = 1$. Assuming zero-mean, stationary, Gaussian processes with correlation matrix \cite{18}
\begin{equation}
  \begin{aligned}
      C_{ij}(\tau) &= \langle \xi_i(t+\tau) \xi_j(t) \rangle = \sigma_i \sigma_j r_{ij} e^{-\lvert \tau \rvert / \tau_c} \\
      S_{ij}(\omega) &= \frac{2 \sigma_i \sigma_j r_{ij} \tau_c}{1 + \omega^2 \tau_c^2}
  \end{aligned}
\end{equation}
where $r_{ii} = 1$, $|r_{ij}| \leq 1$. We use the correlation coefficient notation, when two qubits have equal amplitudes, we can write them as $\sigma_1 = \sigma_2 = \sigma$ and $r_{12} = \rho_{12}$.

Since $[\hat{H}(t_1), \hat{H}(t_2)] = 0$, any off-diagonal element $\rho_s(t)$ only acquires a random phase $\phi_{random}(t)$, which is linear in $\epsilon$. The stationary Gaussian noise and the coherence decay is fully determined by its variance $\chi_{\alpha\beta}(t)$. The second-order cumulant holds Kubo relation. \cite{12}
\begin{equation}
  \frac{\rho_{\alpha \beta}(t)}{\rho_{\alpha \beta}(0)} = e^{-\chi_{\alpha \beta}t}, \quad \chi_{\alpha \beta}(t) = \frac{1}{2} \operatorname{Var}(\Phi_{\alpha \beta}(t))
\end{equation}
Letting $\mathbf{s}_{\alpha\beta}$ collect the $\hat{\sigma}_z$ eigenvalue differences by the pair $(\alpha,\beta)$. The accumulated stochastic phase associated with the coherence $\rho_{\alpha\beta}(t)$ and 
$\Phi_{\alpha\beta}(t) = \int_0^t \mathbf{s}_{\alpha\beta}^{\!\top}\,
\boldsymbol{\xi}(t')\,\mathbf{y}(t')\,\mathrm{d}t'$. The superscript $\top$ denotes matrix transpose. The time-domain forms is \cite{26}
\begin{equation}
  \chi_{\alpha \beta}(t) = \frac{1}{2} \int_0^t \int_0^t s_{\alpha \beta}^\top \left(C(t_1 - t_2) \circ \mathbf{y}(t_1) \mathbf{y}^\top(t_2) \right) s_{\alpha \beta} \, dt_1 dt_2
\end{equation}
Here, the symbol \(\circ\) denotes an Hadamard product. For the Bell coherence between $\{|01\rangle,|10\rangle\}$, one has $\mathbf{s}_{01,10}=(+1,-1)^{\!\top}$, and write out $\mathbf{y}(t)=(y_1(t),y_2(t))^{\!\top}$.
The frequency-domain forms becomes \cite{26}
\begin{align}
  \chi_{\alpha \beta}(t)
  &= \frac{1}{\pi} \int_{0}^{\infty} 
      \mathbf{s}_{\alpha \beta}^{\top}
      \left(S(\omega) \circ \mathbf{Y}(\omega, t) \mathbf{Y}^{\top}(\omega, t)\right)
      \mathbf{s}_{\alpha \beta}\, \mathrm{d}\omega \\
  Y_i(\omega, t)
  &= \int_{0}^{t} y_i(t') e^{i\omega t'}\, \mathrm{d}t'
\end{align}

For the no-DD baseline, we take \(y_i(t) = 1\).
\begin{equation}
 |Y_i(\omega,t)|^2 = F(\omega,t)=\frac{\sin^2(\omega t/2)}{(\omega/2)^2}
\end{equation}
 Using the Ornstein-Uhlenbeck (OU) \cite{27,28} kernel to evaluating Eqs.(5)-(8), and yields the single-qubit dephasing exponent. The non-Markovian fluctuations of low-frequency OU noise models naturally arise from ensembles of two-level systems (TLSs). As shown in Fig.2(a), compared with the Markovian limit, OU-type non-Markovian memory yields a markedly slower, non-exponential decay in fidelity of $\ket{\hat{\Phi}^+}$. We define $\lambda_i^2 = 2\sigma_i^2$. 
\begin{equation}
\chi^{(1)}_i(t)=\lambda_i^2 \tau_c\!\left(t-\tau_c\!\left(1-e^{-t/\tau_c}\right)\right)
\end{equation}
Here, $\chi^{(1)}_i(t)$ denotes the single-qubit dephasing exponent produced by the OU noise acting independently on qubit $i$ (See Supplemtantal Material). The two-qubit exponent follows from Eq.(5) as
\begin{equation}
\chi_{01,10}(t)=\chi^{(1)}_1(t)+\chi^{(1)}_2(t)-2\,\rho_{12}\,\sqrt{\chi^{(1)}_1(t)\,\chi^{(1)}_2(t)}
\end{equation}
In the symmetric case, we take $\lambda_1=\lambda_2=\lambda$ and $\rho_{12}=\rho$.
\begin{equation}
\chi_{01,10}(t)=2(1-\rho)\,\chi^{(1)}(t),
\chi^{(1)}= \chi^{(1)}_1=\chi^{(1)}_2
\end{equation}

The Bell based $|\hat{\Psi}^ + \rangle = (|01\rangle + |10\rangle)/\sqrt{2}$ under pure dephasing with $\rho_{01,10}(t) = e^{-\chi_{01,10}t}/2$ and unchanged populations. The Wootters concurrence and the fidelity of $|\hat{\Psi}^{+}\rangle$ are \cite{29}
\begin{equation}
C_{noc}(t)=e^{-\chi_{01,10}t},
\mathcal{F}_{\hat{\Psi}^{+}}(t)
= \frac{1 + e^{-\chi_{01,10}t}}{2}.
\end{equation}
A general $X$ state with populations $p_{00},p_{11}$, one may use
$C_{noc}(t)=\max\{0,\,2|\rho_{01,10}(t)|-2\sqrt{p_{00}p_{11}}\}$. Eq.(12) is the special case of $p_{00}=p_{11}=0$.

The correlation function defined in Eq.(3) depends only on the time difference $\tau$. We set $\tau = t$ and writing
$C_{noc}(t)=C_{ij}(\tau = t)$. As shown in Fig.2(b), the entanglement between the two fluxonium qubits quantified by the concurrence $C_{noc}(t)$. As the inter-qubit noise correlation is independent with baths ($\rho\!=\!0$), decay attains to double. As the inter-qubit noise correlation $\rho$ increases, decays more slowly. Increasing $\rho$ enhances the common-mode component of TLS noise, suppresses relative phase fluctuations, slows the decay of the concurrence. As the inter-qubit noise correlation $\rho\!\to\!1$, coherence disappears. Short- and long-time limits come from Eq.(9).
\begin{align}
t\ll\tau_c:\quad & \chi^{(1)}_i(t)=\tfrac{1}{2}\lambda_i^2 t^2+\mathcal{O}(t^3),\\
t\gg\tau_c:\quad & \chi^{(1)}_i(t)\approx \Gamma_{\phi,i}\, t,\Gamma_{\phi,i}=\lambda_i^2\tau_c=S_{ii}(0)
\end{align}
which saturates to $\Gamma_{\phi,1}+\Gamma_{\phi,2}-2 \rho_{12}\sqrt{\Gamma_{\phi,1}\Gamma_{\phi,2}}$ for $t\gg\tau_c$. Hence
\begin{equation}
C_{noc}(t)\approx e^{-\big(\Gamma_{\phi,1}+\Gamma_{\phi,2}-2 \rho_{12}\sqrt{\Gamma_{\phi,1}\Gamma_{\phi,2}}\big)t}
\end{equation}
The instantaneous two-qubit dephasing rate is
\begin{equation}
    \begin{aligned}
    \gamma_{\mathrm{inst}}(t) &= \dot{\chi}_{01,10}(t) \\
    &= \sum_{i=1}^{2}\lambda_i^2 \tau_c\!\left(1-e^{-t/\tau_c}\right)
       -2 \rho_{12}\lambda_1\lambda_2 \tau_c\!\left(1-e^{-t/\tau_c}\right)
    \end{aligned}
    \end{equation}
In the symmetric case, Eq.(16) reduces to $2(1-\rho)\lambda^2\tau_c(1-e^{-t/\tau_c})$. Fig.2(c) shows that the instantaneous effective dephasing rate $\gamma_{eff}(t)=-\, d\ln F/dt$ rises from near zero and saturates around \(t\!\sim\!\tau_c\), delineating the crossover from the quantum-Zeno regime to the Markovian limit.

Comparison against the no-DD baseline, we want to analysis DD baseline, we replace $y_i(t) = 1$ in Eqs.(5)-(7) by the chosen pulse sequence, and use a new sequence filters $Y_i(\omega,t)$ in Eq.(8). The remaining pipeline from $\chi$ to $C_{noc}(t)$ and $\mathcal{F}_{\hat{\Psi}^+}(t)$ is unchanged.

We rephrase OU noise model in terms of a post-Markovian master equation (PMME) for the reduced density matrix of the two fluxonium qubits. In the PMME framework, the non-Markovian dynamics is generated by a time-local Markovian Lindbladian $\mathcal{L}$ dressed by a memory kernel $k(t)$.
\begin{equation}
	\frac{d}{dt}\hat{\rho}(t)
	= \mathcal{L}\int_0^t k(t-\tau)\, e^{\mathcal{L}(t-\tau)} \hat{\rho}(\tau)\, d\tau .
\end{equation}
Here, $\rho(t)$ is the reduced density matrix of the two fluxonium qubits, $\mathcal{L}$ is a standard Markovian generator, and $k(t)$ encodes the memory of the bath. Non-Markovian effects are incorporated through the scalar memory kernel $k(t)$. 

When pure dephasing of two fluxonium qubits coupled to a correlated TLS environment, we choose the Markovian generator
\begin{equation}
	\mathcal{L}\rho
	= \frac{\gamma_0}{2}\sum_{i,j=1}^2 \hat{R}_{ij}
	\left(
	\hat{\sigma}_z^{(i)} \rho \hat{\sigma}_z^{(j)}
	- \frac{1}{2}\{\hat{\sigma}_z^{(i)}\hat{\sigma}_z^{(j)}, \rho\}
	\right),
\end{equation}
where $\hat{R}_{ij}=\hat{\rho}_{ij}$ sets the overall dephasing scale. Eq.(18) describes correlated Markovian dephasing generated by the TLS bath in the limit of vanishing memory.

Motivated by the exponential OU correlation function and the corresponding Lorentzian spectral density of the TLS bath, we take
\begin{equation}
	k(t) = \frac{1}{\tau_c} e^{-t/\tau_c}\,\Theta(t),
\end{equation}
where $\tau_c$ is the bath correlation time and $\Theta(t)$ is the Heaviside step function. The PMME dynamics reproduces the OU correlation function and the Lorentzian spectrum used in our classical-noise description.

In particular, the off-diagonal matrix elements of $\rho(t)$ obey
\begin{equation}
	\hat{\rho}_{m,n}(t) = e^{-\chi_{\mathrm{PM}}t} \hat{\rho}_{m,n}(0),
\end{equation}
with a non-Markovian dephasing functional $\chi_{\mathrm{PM}}(t)$ determined by the kernel $k(t)$ and the generator $\mathcal{L}$. In the Markovian limit $\tau_c \rightarrow 0$, the kernel becomes an sharply peak and $\chi_{\mathrm{PM}}(t)$ reduces to linear form $\chi_{\mathrm{M}}(t) = \Gamma_{\varphi} t$ with a constant dephasing rate $\Gamma_{\varphi}$. For finite $\tau_c$, $\chi_{\mathrm{PM}}(t)$ coincides with the filter-function result $\chi(t)$ obtained from the OU noise model, and the corresponding instantaneous dephasing rate becomes strongly time dependent.
\begin{equation}
	\gamma_{\mathrm{inst}}(t) = \frac{d}{dt}\chi_{\mathrm{PM}}(t)
\end{equation}
This time-dependent rate provides a convenient quantitative link between the classical OU picture and a fully quantum non-Markovian master-equation description, and it naturally captures Zeno-like plateaus and partial revivals of coherence observed in the concurrence dynamics.

We include dynamical decoupling (DD) into the noise analysis, we start from the pure dephasing Hamiltonian in Eq.(2) and move to the modulation frame defined by the control Hamiltonian $\hat{H}_c(t)$ that generates the sequence of $\pi$ pulses. The control propagator is $\hat{U}_c(t)=\mathcal{T}e^{-i\int_0^t \hat{H}_c(t')\,dt'}$, in the modulation frame, we have $\hat{U}_c^\dagger(t) \hat{H}(t) \hat{U}_c(t)$. For a sequence of instantaneous $\pi$ pulses applied to the qubits, the longitudinal TLS noise retains its diagonal $\hat{\sigma}_z$ structure but acquires a time dependent modulation. $y_i(t)=\pm 1$ is the modulation function that flips sign at each $\pi$ pulse in Eq.(2). In the presence of DD, the nontrivial time dependence of $y_i(t)$ encodes the effect of the pulse sequence. We leave the noise spectrum unchanged, the filter function associated with the sequence is \cite{26}
\begin{equation}
	\begin{aligned}
    Y_i(\omega,T)&=\int_0^T y_i(t') e^{i\omega t'} dt',\\ 
       F_{ij}(\omega,T)&=Y_i(\omega,T)Y_j^\ast(\omega,T).
   \end{aligned}
\end{equation}
Comparing with Eq.(3), the OU spectral density becomes
\begin{equation}
S_{ij}(\omega)=\frac{2\sigma_i\sigma_j \hat{\rho}_{ij}\tau_c}{1+\omega^2\tau_c^2},
\end{equation}
the dephasing exponent for any two fluxonium qubits coherence becomes
\begin{equation}
\chi_{\alpha\beta}^{\rm DD}(T)
=\frac{1}{\pi}\int_0^\infty
s_{\alpha\beta}^{\top}\!\left(S(\omega)\circ F(\omega,T)\right)
s_{\alpha\beta}\, d\omega,
\end{equation}
The DD generalization of the free-evolution result.
For the Bell coherence $\{|01\rangle,|10\rangle\}$, we have
\begin{equation}
	\begin{aligned}
        C_{noc}^{\rm DD}(T)&=e^{-\chi_{01,10}^{\rm DD}T},
        F_{\hat{\Psi}^+}^{\rm DD}(T)&=\frac{1+e^{-\chi_{01,10}^{\rm DD}T}}{2}.
    \end{aligned}
\end{equation}
The standard sequences such as CPMG, UDD, and XY-8 mainly target either high-frequency noise or pulse-error compensation. In contrast, our aim is to design a two fluxonium qubits DD pattern capable of suppression correlated longitudinal TLS noise. 

\begin{figure}[h!]
	\centering
\includegraphics[width=1.1\linewidth]{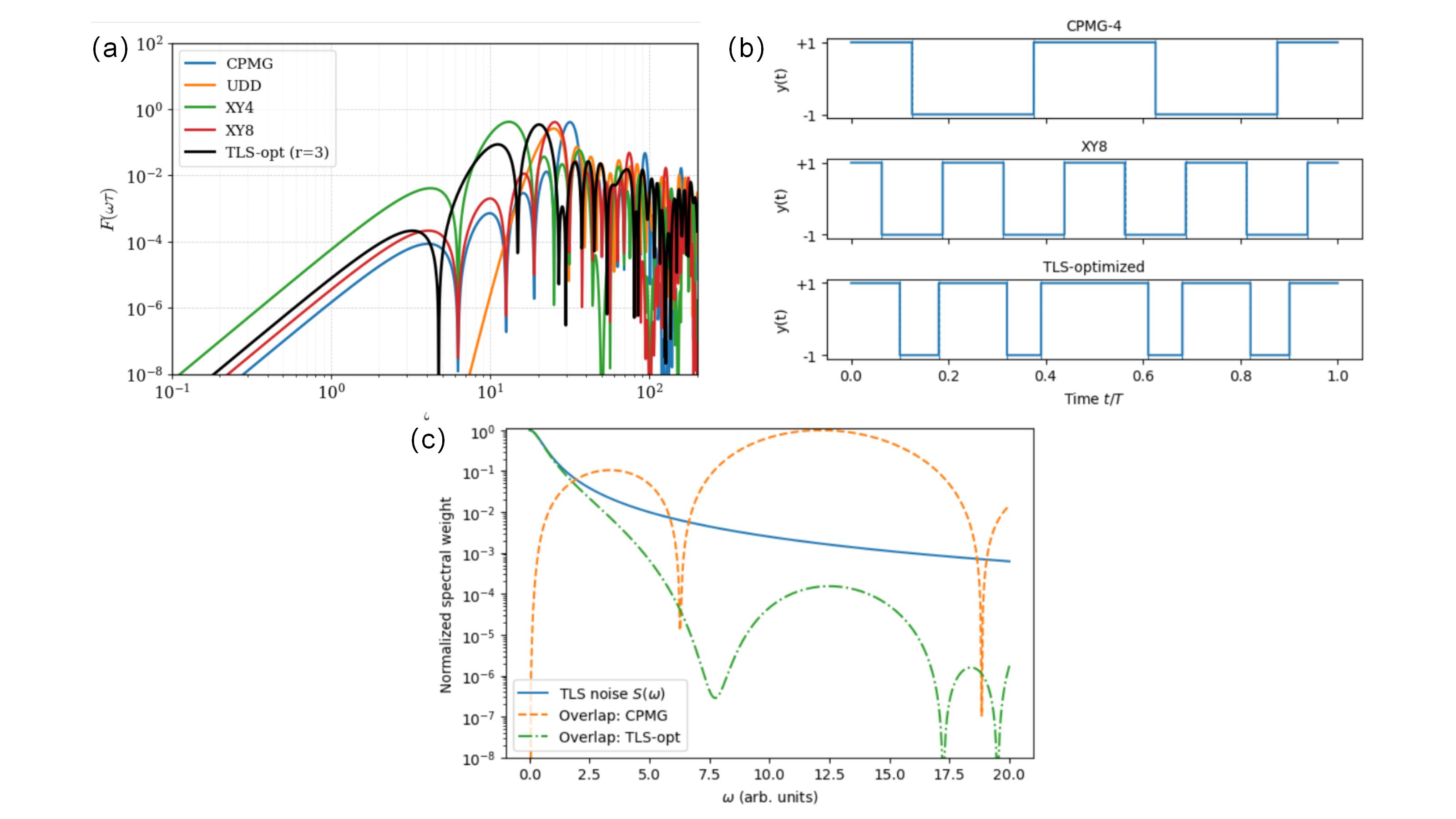}
	\caption{(a) Comparison of filter functions $F(\omega,T)$ for several DD sequences, showing that the TLS-opt sequence provides the deepest suppression at low frequencies. (b) Overlap between the TLS noise spectrum $S(\omega)$ and the filter functions, where the TLS-opt sequence strongly reduces low-frequency contributions. (c) Time-domain modulation functions $y(t)$ for CPMG-4, XY-8, and TLS-opt. 
	}
\end{figure}
As shown in Fig.3(a), different DD sequences exhibit distinct filter functions. CPMG, UDD and XY-8 sequences generate suppression notches at different frequencies, while the TLS-optimized sequence produces significantly deeper low-frequency suppression consistent with TLS noise characteristics. Fig.3(a) compares the filter functions of the no-DD evolution, CPMG, XY-8, and TLS-opt sequences for the same $N$ and $T^*$. While CPMG and XY-8 generate broad high-frequency notches around $\omega T^* \sim N\pi$, the TLS-opt sequence produces a much deeper suppression in the low-frequency region $\omega \tau_c \ll 1$, where the TLS noise is strongest.

The DD sequence is based on the two fluxonium qubits Heisenberg-Weyl group, which applied as a symmetrized cycle.
\begin{equation}
\mathcal{G}=\{\hat{I},\hat{X},\hat{Y},\hat{Z}\}^{\otimes 2}
\end{equation}
 A cycle consists of an order of two fluxonium qubits $\pi$ pulses that effectively symmetrizes the longitudinal noise
operators. In the leading-order Magnus expansion, this cycle averages the operators $\hat{\sigma}_z^{(1)}=\hat{\sigma}_z^{(2)}=0$, which cancel
the dominant TLS-induced dephasing.

TLS-induced frequency fluctuations follow an OU process with a Lorentzian spectrum
\begin{equation}
S_{\rm TLS}(\omega)\propto \frac{1}{1+\omega^2\tau_c^2}
\end{equation}
whose weight is strongly concentrated in the low-frequency regime
$|\omega|\lesssim 1/\tau_c$. Standard DD sequences place their strongest
suppression notch near frequencies $\omega\sim N\pi/T$, which is generally too
high to efficiently suppress the Lorentzian peak.
Here, the integer $N$ denotes the total number of $\pi$ pulses applied within the storage interval $T$. Increasing $N$ refines the time resolution of the modulation function $y(t)$ and shifts the main suppression notch of the filter function toward higher frequencies. The optimization leads to sequence spectral match to the Lorentzian OU noise, we name it TLS-opt. Fig.3(b) illustrates the modulation functions $y(t)$ for three DD sequences: CPMG-4, XY-8, and TLS-opt. The first two sequences feature uniformly spaced $\pi$ pulses with different rotation axes, while the TLS-opt sequence consists of nonuniform pulse locations chosen to satisfy $\langle y(t)\rangle_T = \int_0^T y(t)dt/T = 0$, which produces a pronounced low-frequency notch tailored to the Lorentzian TLS noise spectrum.

The pulse locations within a total storage time $T$ are denoted $t_k$ and will be optimized to match the TLS spectrum. To tailor the sequence to the TLS spectrum, for a target storage time $T^\ast$, we choose the pulse times
$t_k$ to minimize the overlap integral
\begin{equation}
\begin{aligned}
		&\mathcal{J}[t_k]= \chi_{0,1}^{\mathrm{DD}}(T^{\ast}) \\
		&= \frac{1}{\pi}\int_0^{\infty} S_{\mathrm{TLS}}(\omega)\times F(\omega,T^{\ast};t_k )d\omega,
\end{aligned}
\end{equation}
where $S_{\mathrm{TLS}}(\omega)$ is the Lorentzian spectrum. Minimizing Eq.(28) forces the filter function to develop a deep suppression notch at $\omega\simeq 0$, where the TLS noise is strongest. Fig.3(c) displays the modulation functions $y_i(t)$ for the two fluxonium qubits for CPMG, XY-8, and TLS-opt. The spectrum overlap between the Lorentzian TLS noise $S_{\mathrm{TLS}}(\omega)$ and the filter functions of different DD sequences. CPMG exhibits a relatively large overlap in the low-frequency region, whereas the TLS-opt protocol suppresses the overlap by several orders of magnitude. This demonstrates that the optimized pulse timings can minimize the cost functional $\mathcal{J}$ and achieve true spectrum match to the TLS environment.

Once a DD sequence is specified, the concurrence and fidelity follow directly from the modified dephasing exponent. The corresponding time scales include: the entanglement lifetime $\tau_C$ defined by $C(\tau_C)=1/e$, the high-fidelity time $T_{0.999}$ defined by $\mathcal{F}_{\hat{\Psi}^+}(T_{0.999})=0.999$.

Because the TLS-optimized sequence strongly suppresses the low frequency component of $S_{\rm TLS}(\omega)$, the effective dephasing strength is reduced by a large factor, which leads to a much slower decay of concurrence and fidelity.
\begin{equation}
\chi_{01,10}^{\rm TLS\text{-}opt}(T)
\ll \chi_{01,10}^{\rm CPMG}(T)
\ll \chi_{01,10}^{\rm no-DD}(T),
\end{equation}
 We can present the time evolution of the concurrence under different DD sequences. The TLS-opt protocol yields the slowest decay, extending the entanglement lifetime by nearly an order of magnitude compared with CPMG, XY-8, and the no-DD baseline. This behavior is fully consistent with the spectrum overlap trends shown in Fig.3(b). 
In the long-time limit, $T\gg \tau_c$, the no-DD exponent grows approximately linearly, $\chi^{\rm no-DD}(T)\propto T$. Standard DD sequences reduce the slope modestly. In contrast, the TLS-opt sequence reduces the effective low-frequency noise amplitude by a factor of $3-10$.

We present a comparison of $C(T)$ and $\mathcal{F}_{\hat{\Psi}^+}(T)$ for four cases: no-DD, CPMG, XY-8, and the TLS-opt protocol. The optimized sequence exhibits the slowest decay and the highest stored fidelity within the micro-second regime, confirming its advantage for non-Markovian TLS noise.
\nocite{*}

\section{NUMERICAL SIMULATION}

In this section, we present numerical simulations of the two fluxonium qubits under TLS-induced non-Markovian dephasing and different dynamical decoupling protocols. We first specify the parameter set and connection to existing fluxonium experiments, then discuss the optimization of TLS-opt sequences, and finally compare the entanglement dynamics and robustness of different protocols.

We consider two capacitively coupled fluxonium qubits biased at their respective sweet spots and operated without micro-wave drive (see Supplementary materials). The bare qubit frequencies and coupling strength are chosen as $\omega_{q_1}/2\pi = 0.6GHz$, $\omega_{q_2}/2\pi = 0.62GHz$, and $\mathcal{J}/2\pi = 0.02GHz$.

The correlated TLS is modeled by the OU noise process with correlation time $\tau_c$ and amplitude $\lambda$, the Lorentzian spectral density is introduced. For concreteness, we take $\tau_{c} = 0.5\mu s$ and $\lambda/2\pi = 80KHz$, 
which is compatible with a TLS density of order $10^{18}m^{-3}$ and individual qubit-TLS couplings in the $10-100KHz$ range. The dimensionless correlation coefficient $\rho$ is used to interpolate between independent and fully correlated noise sources, we choose $\rho=0.8$ to emphasize the role of collective dephasing on entanglement.

The concurrence of Bell based $\ket{\hat{\Psi}^+}=(\ket{01}+\ket{10})/\sqrt{2}$ exhibits pronounced non-exponential decay and partial revivals in no-DD. The instantaneous dephasing rate $\gamma_{inst}(t)$, obtained either from the OU filter-function formalism or from the PMME description, shows a rapid initial rise followed by a slow approach to its Markovian plateau, reflecting Zeno-like protection at short times.

We now specify the optimization procedure used to construct TLS-opt DD sequences within the Heisenberg-Weyl group $\mathcal{G}^{\otimes 2}$. For a given total storage time $T^*$ and a fixed number $N$ of $\pi$ pulses, the control sequence is characterized by the normalized pulse times
\begin{equation}
	t_k/T^* = \{t_1/T^*,\,t_2/T^*,\,\dots,\,t_N/T^*\},
\end{equation}
with $0 < t_1 < t_2 < \cdots < t_N < T^*$. In practice, we perform the optimization numerically by using the Nelder-Mead simplex algorithm, starting from an equally spaced initial guess $t_k{(0)}/T^* = (k-1/2)/N$. The pulse times are constrained within the open interval $(0,T^*)$ and preserve their ordering. For $N=8$ and $T^*=1\mu s$, a typical optimized solution takes the form $t_k/T^{*} \approx \{0.08,\,0.19,\,0.31,\,0.43,\,0.57,\,0.70,\,0.82,\,0.92\}$, which is notably skewed toward the beginning and end of the interval, reflecting the time dependence of $\gamma_{\mathrm{inst}}(t)$ and the enhanced sensitivity to low-frequency components of the Lorentzian spectrum.

The minimized cost function is the weighted overlap between the TLS noise spectrum and the filter function of the sequence.
\begin{equation}
	J[t]=
	\frac{1}{\pi}
	\int_{0}^{\infty} S_{\mathrm{TLS}}(\omega)\,\times
	F(\omega,T^*;t)\, d\omega,
\end{equation}
where $F(\omega,T^*,t\}$ is the filter function defined in Eq.(28).

The OU parameters are the same as in Table I, and the DD pulses are assumed to be ideal, instantaneous $\pi$ rotations.
\begin{table}[hptp]  
	\caption{Parameters for two fluxonium qubits and TLS environment.}    
		\begin{tabular}{|c |c |c|}  
			\hline
			Quantity & Symbol & Value / Origin \\ \hline  
			Qubit frequencies & $\omega_{q}^{(1)}, \omega_{q}^{(2)}$ & $0.60, 0.62GHz$ \\ \hline  
			Exchange coupling & $J$ & $0.02GHz$ (capacitive) \\ \hline  
			Bath correlation time & $\tau_c$ & $0.5\ \mu s$ (OU noise) \\ \hline  
			Noise amplitude & $\lambda/{2\pi}$ & $80KHz$ (TLS couplings) \\ \hline  
			Correlation coefficient & $\rho$ & $0.8$(common-mode)\\ \hline  
			TLS density & - & $\sim 10^{18} m^{-3}$ (materials) \\ \hline  
			Storage time window & $T^{*}$ & $1\ \mu s$ (NISQ gate times) \\ \hline 
		\end{tabular}  
\end{table}  
 
 We now turn to the entanglement dynamics under the different DD protocols. 
 \begin{table*}
	\caption{Entangled time extracted from two-fluxonium setting.}
\renewcommand{\arraystretch}{1.15}
    \label{}
	\centering
    \begin{tabular}{|c|c|c|c|c|}
		\hline
		Protocol &
		$\tau_C^{no-DD} (\mu s)$ &
		$\tau_C^{TLS-opt} (\mu s)$ &
		$T_{0.999}^{no-DD} (\mu s)$ &
		$T_{0.999}^{TLS-opt} (\mu s)$ \\
		\hline
		Two-fluxonium 
		correlated TLS noise& 0.15 & 1.3 & 0.02 & 0.35\\	
		\hline
	\end{tabular}
\end{table*}
 In the no-DD case, the concurrence decays rapidly and exhibits only small revivals due to the partial information backflow from the TLS bath. The entanglement lifetime $\tau_C$ defined as the time when $C(T)$ first drops below $1/e$, $\tau_C^{no-DD}\approx 0.15 \mu s$. The CPMG sequence substantially slows down the decay, extending the lifetime to $\tau_C^{CPMG}\approx 0.45\mu s$, while XY-8 yields a further improvement to $\tau_C^{XY-8}\approx 0.7\mu s$. The TLS-opt protocol achieves the strongest protection: the initial decay becomes nearly flat on the semi-logarithmic scale, and the concurrence remains above $1/e$ up to $\tau_C^{TLS-opt}\approx 1.3\mu s$, corresponding to an enhancement by almost one order of magnitude compared with free evolution. A similar trend can be obtained if we use the time $T_{0.999}$ at which the Bell-based fidelity $\mathcal{F}(T)$ drops below $0.999$.  To further quantify the practical impact of our TLS-tailored control on multi-qubits performance. Table II summarizes the entanglement from our simulations. We list the concurrence coherence time $\tau_C$ and the high-fidelity operating window $T_{0.999}$ (the Bell-based fidelity satisfies $\mathcal{F}\ge 0.999$), for both free evolution (no-DD) and the TLS-opt DD protocol. The resulting gains, $\tau_C^{\mathrm{TLS-opt}}/\tau_C^{\mathrm{no-DD}}\approx 8.7$ and $T_{0.999}^{\mathrm{TLS-opt}}/T_{0.999}^{\mathrm{no-DD}}\approx 17.5$, highlight that our spectrum matched sequence substantially enlarges the error relevant to near term error mitigation and repeated entangling gate cycles.

\begin{figure}[hptp]
	\centering
\includegraphics[width=1\linewidth]{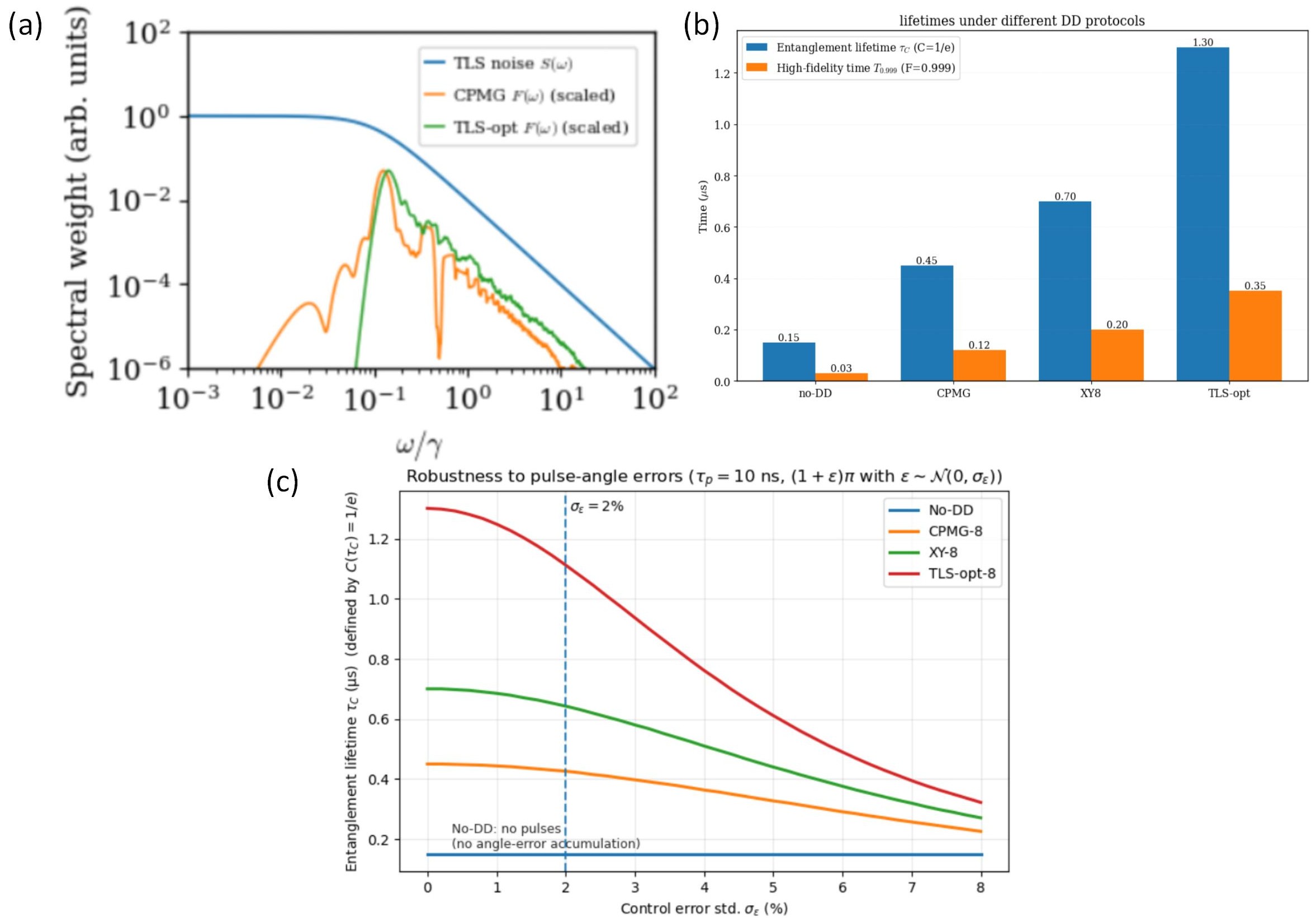}
	\caption{(a) Comparison between the TLS noise spectrum $S(\omega)$ and the (scaled) filter functions $F(\omega)$, showing deeper low-frequency suppression and reduced spectral overlap for the TLS-opt sequence. (b) Summary of performance under different DD protocols: entanglement lifetime $\tau_C$ defined by $C=1/e$ and high-fidelity time $T_{0.999}$ defined by $\mathcal{F}=0.999$. (c) Robustness to pulse-angle errors: $\tau_C$ versus the control-error standard deviation $\sigma_\epsilon$, where each $\pi$ pulse is implemented as $(1+\epsilon)\pi$ with $\epsilon\sim\mathcal{N}(0,\sigma_\epsilon)$ and $\tau_p=10ns$.
	}
\end{figure}
As shown in Fig.4(a), the TLS noise spectrum is dominated by low-frequency components, and the TLS-opt sequence reshapes the filter function to achieve deeper suppression in this frequency range. Consequently, the spectral overlap between $S(\omega)$ and $F(\omega)$ is reduced compared with standard sequences, providing the physical mechanism behind the improved coherence/entanglement preservation. Fig.4(b) summarizes the key performance metrics across different DD protocols: both the entanglement lifetime $\tau_C$ and the high fidelity time $T_{0.999}$ are enhanced by dynamical decoupling, with the TLS-opt protocol achieving the largest improvement. This compact comparison highlights that the advantage of TLS-opt is consistent for both entanglement-based and fidelity-based benchmarks. To assess experimental robustness, Fig.4(c) shows $\tau_C$ as a function of the pulse-angle error level $\sigma_\epsilon$: increasing control errors gradually reduce the benefit of all pulsed protocols, while the no-DD baseline remains essentially unchanged due to the absence of pulse accumulation. Importantly, TLS-opt retains a clear advantage over CPMG-8 and XY-8 for moderate $\sigma_\epsilon$, indicating that the optimization is not restricted to the ideal-control limit.

These results confirm that tailoring the DD filter function to the Lorentzian TLS spectrum provides a substantial gain over standard periodic sequences that are optimized for broadband or $1/\omega$-type noise. Importantly, the non-Markovian character of the bath is crucial: if we artificially reduce the correlation time to $\tau_c\ll 1/\lambda$, such that the OU process approaches the Markov limit, the advantage of TLS-opt shrinks and all protocols converge to similar lifetimes determined by a constant dephasing rate. By comparing the two contributions, $\chi_{tot}=\chi_{TLS}+\chi_{ctrl}$: once $\chi_{TLS}$ is filtered below $\chi_{ctrl}$, different protocols exhibit similar performance.

In realistic devices, DD pulses have finite duration and are subject to control errors. To assess the robustness of the TLS-opt protocol, we model each $\pi$ pulse as a square pulse of duration $\tau_p=10ns$ with a small relative rotation angle error $\epsilon$, such that the actual rotation angle is $(1+\epsilon)\pi$. We average the concurrence dynamics over random realizations of $\epsilon$ drawn from a zero-mean Gaussian distribution with standard deviation $\sigma_\epsilon$.

For moderate control errors ($\sigma_\epsilon \lesssim 2\%$), the relative ordering of the protocols remains unchanged: TLS-opt provides the longest entanglement lifetime, followed by XY-8 and CPMG. As $\sigma_\epsilon$ increases, pulse imperfections introduce additional errors that accumulate with the number of applied $\pi$ pulses (roughly scaling with $N$) and progressively reduce the contrast between different sequences. Beyond a certain $\sigma_\epsilon$ (for a given $N$), the improvement from further suppressing TLS-induced dephasing becomes marginal, and the observed decay is primarily governed by control-induced errors rather than by the residual TLS contribution. 

\begin{table*} 
	\caption{Comparison between our theory and non-Markovian dynamics relevant to superconducting qubits experimental results.}
\begin{tabular}{c c c}
	\hline
Platform & Mechanism & Key parameters\\
	\hline
\makecell*{IBM-Q transmon \\(single-qubit\\idle, ibmq\_athens) \cite{30}} & \makecell*{Post-Markovian Master Equation \\(PMME) with memory kernel, coherent\\crosstalk, incoherent noise} &\makecell*{Testing-set trace-norm distance: \\Markovian model median:
	\\0.12 (95\%: 0.18) \\PMME median: 0.06 (95\%: 0.08)} \\
	\hline
\makecell*{ Transmon idle evolution\\(temporally correlated noise\\	model with\\
	superconducting-qubit) \cite{31}} & \makecell*{Temporally correlated dephasing with\\ finite-bandwidth spectrum} &\makecell*{Parameter: $1/f^{\alpha}$ \\with $\alpha=0.95$ and cutoff
	$\omega_c=3\omega_q$ \\gives an effective $T_2 \approx 10\mu$s \\
	Ohmic-like case: $\omega_c=5\omega_q$.} \\
		\hline
	\makecell*{ Fluxonium \cite{32}} & \makecell*{ Non-Markovian relaxation spectrum\\dominated by discrete TLS resonances} &\makecell*{TLS band: $0.1-0.4GHz$ \\TLS lifetimes: $ms$\\
		The inferred dipole moment: $\sim 6$ Debye\\Couplings:
		$g/2\pi \approx 4.9-88KHz$ \\Mean area density:
		$\sim 0.4GHz m^{-2}$} \\
		\hline
	\makecell*{Multi-qubit \\superconducting devices  \cite{33}} & \makecell*{Non-Markovian process
		reconstruction\\(beyond Markovian assumptions)} &\makecell*{Reconstruction predicts\\ device behavior with\\ infidelity at
		$\sim 10^{-3}$ level.} \\
        \hline
	\makecell*{Two fluxonium qubits\\ correlated TLS noise\\ (Our work)} & \makecell*{PMME-consistent colored-noise TLS model\\ (OU) + TLS-optimized DD} &\makecell*{$\rho=0.8,J=0.02GHz$ \\ $\lambda/2\pi=80KHz$\\
		TLS density $\sim 10^{18}{m^{-3}}$} \\
	\hline
\end{tabular}
\end{table*}

Table III summarizes research on non-Markovian mechanisms in superconducting-qubit(-like) platforms. The table highlights the platform, the dominant mechanism that gives rise to memory effects and the key parameters. Compared with studies centered on single-qubit spectroscopy or general process, our work directly targets multi-qubits and
employs a spectrum-matched dynamical-decoupling design to reduce the effective overlap between the Lorentzian
TLS spectrum and the control filter function. 

\section{CONCLUSION}
We develope a compact framework to analyze and suppress correlated non-Markovian dephasing in a two-Fluxonium register dominated by an OU TLS bath, and we connect the resulting memory effects to entanglement (concurrence) and Bell-based fidelity with a PMME-consistent interpretation. Building on a two-qubit Heisenberg-Weyl decoupling cycle, we optimize the pulse locations to minimize the spectral-overlap cost with the Lorentzian TLS spectrum, yielding a TLS-tailored protocol with markedly deeper low-frequency suppression than standard periodic sequences.  By including finite pulse duration and rotation-angle errors, we find that TLS-opt remains advantageous under moderate imperfections. The key requirement for exploiting TLS-opt DD is not an extremely high pulse count, but the ability to implement $\sim 10$ well-calibrated high fidelity refocusing $\pi$ pulses within micro-second region. TLS-opt DD extends the useful entanglement window to time scales compatible with two-qubit gate durations and readout times in state-of-the-art fluxonium devices, thereby offering a viable route toward higher-fidelity entangling operations in NISQ-era processors.

\begin{acknowledgments}
This work was funded by the State Key Laboratory of Quantum Optics Technologies and Devices, Shanxi University, Shanxi, China (Grants No.KF202503).
\end{acknowledgments}

\bibliography{ref}

\end{document}